\documentclass[ifip]{svmult}
\usepackage[latin1]{inputenc}
\usepackage[T1]{fontenc}
\usepackage{textcomp}
\usepackage{xspace}
\usepackage[english]{babel}
\usepackage[bottom]{footmisc}
\usepackage{amsmath,amssymb}
\usepackage{tabularx}
\usepackage[hang]{subfigure} 
\providecommand{\href}[2]{#2}

\providecommand{\hypersetup}[1]{}
\providecommand{\url}[1]{#1}

  \usepackage[dvips]{graphicx}        

\newcommand{\ie}{, {i.e.},\xspace}
\newcommand{\eg}{, {e.g.},\xspace}

\newcommand{\DEF}{
  \mathbin{\smash[t]{\overset{\scriptscriptstyle\mathrm{def}}{=}}}}

\newcommand{\NN}{{\ensuremath{\mathord{\mathbb{N}}}}\xspace}

\newcommand{\CA}{\ensuremath{A}\xspace}
\newcommand{\CB}{\ensuremath{B}\xspace}
\newcommand{\VS}{\textsf{VSec}\xspace}
\newcommand{\Arc}{\textsf{Arc}\xspace}
\newcommand{\TTA}{\textsf{T1}\xspace}
\newcommand{\TTB}{\textsf{T2}\xspace}
  
\begin{document}

\title*{Non-Repudiation in Internet Telephony\vspace*{-2mm}}
\author{Nicolai Kuntze\inst{1} \and Andreas U. Schmidt\inst{1}\and
Christian Hett\inst{2}\vspace*{-1mm}}
\institute{Fraunhofer--Institute for Secure Information Technology SIT\\
Rheinstraße 75, 64295 Darmstadt, Germany\\
\texttt{\{andreas.u.schmidt,nicolai.kuntze\}@sit.fraunhofer.de}
\and ARTEC Computer GmbH\\ Robert-Bosch Straße 38, 61184 Karben, Germany\\
\texttt{christian.hett@artec-it.de}}
\maketitle
\begin{abstract}
\vspace*{-2mm}
We present a concept to achieve non-repudiation for natural language conversations over the Internet.
The method rests on chained electronic signatures applied to pieces of packet-based, 
digital, voice communication. 
It establishes the integrity and authenticity of the bidirectional data stream 
and its temporal sequence and thus the security context of a conversation. 
The concept is close to the protocols for Voice over the Internet (VoIP),
provides a high level of inherent security, and extends naturally to multilateral
non-repudiation\eg for conferences. 
Signatures over conversations can become true 
declarations of will in analogy to electronically signed, 
digital documents.
This enables binding verbal contracts, 
in principle between unacquainted speakers, and in particular without witnesses. 
A reference implementation of a secure VoIP archive is exhibited.
\vspace*{-4mm}
\end{abstract}
\section{Introduction}\label{intro}
\vspace*{-2mm}
The latest successful example for the ever ongoing convergence of 
information technologies is Internet based telephony, transporting voice over the Internet protocol (VoIP). 
Analysts estimate a rate of growth in a range of 20\% to 45\% annually, 
expecting that VoIP will carry more than fifty percent of business voice traffic (UK) in a few years~\cite{VoIP-Study}. 
The success of VoIP will not be limited to cable networks, convergent speech and data transmission will 
affect next generation mobile networks as well. 
The new technology raises some security issues. 
For eavesdropping traditional, switched analogue or digital phone calls, 
an attacker needs physical access to the transport medium. 
Digital networks are generally more amenable to attacks, as 
holds already for ISDN and to a yet greater extent for IP networks. 
Efforts to add security features to VoIP products are generally insufficient, 
though proposals exist for the protection of confidentiality and privacy. 
Secure VoIP protocols, using cryptographic protection of a call, 
would even be at an advantage compared to 
traditional telephony systems. 
Protocols like SRTP~\cite{SRTP} can provide end-to-end security to phone calls, 
independently of the security of transport medium and communication provider.

With VoIP maturing, it becomes natural to ask for application-level security
in the context of IP telephony. 
Our purpose is to achieve non-repudiation in this context\ie
for speech over packet-oriented, digital channels, and in particular for VoIP conversations.
This means the capability to produce tenable evidence that a conversation with the alleged contents
has taken place between two or more parties. 
Ancillary information\eg that the conversation partners have designated, personal identities, and the
time at which the conversation has taken place, may be of utmost importance in this regard, 
either to establish a supporting plausibility\eg `caller was not absent during the alleged call', 
or to express relevant semantic information\eg `telephonic order came in before stock price rose'.   
For electronic documents this kind of non-repudiation is commonly achieved by applying 
electronic signatures based on asymmetric cryptography.
In the communication between several parties, the desired result is a binding contract, 
and in analogy the central goal of the present contribution is
a technology to establish binding verbal contracts without witnesses.

This subject has a long pre-history: As early as 1905, Edison proposed the recording
of voice, which was patented 1911~\cite{Edison1911}. 
With the advent of digital signature technology, Merkle~\cite{Merkle1989} envisioned,
referring to Diffie and Hellman that ``Digital signatures promise to revolutionize business by phone''.
However, work on non-repudiation of digital voice communication is scarce.
The work most closely related to ours is the proposal in~\cite{Strasser-Thesis2001},
resting on the theory of contracts and multi-lateral security~\cite{KKKZ2000}. 
It comprises a trusted third party (`Tele-Witness') that is invoked by communicating 
parties to securely record conversations and make them available as evidence 
at any later point in time.

Non-repudiation of inter-personal communication is interesting
because of its inherent evidentiary value, exposed by
forensic evaluation of the contained biometric data\eg
as an independent means of speaker identification~\cite{noiserobustspeakeraut,fusionspeakeraut}. 
Methods for the latter are advanced~\cite{Hollien2001}, 
yielding to recorded voice a high probative force\eg in a court of law. 
In comparison to other media, specific features of voice 
contribute to non-repudiation. 
Voice communication is interactive~\cite{Goodwin1981} 
and enables partners to make further enquiries in case of 
insufficient understanding. 
This mitigates to some extent problems to which signed 
digital documents are prone\eg 
misinterpretations due to misrepresentation, lack of uniqueness of presentation, 
and inadvertent or malicious hiding of content~\cite{LP98-WYSIWYS}.

We set out requirements for non-repudiation which are very particular
in the case of VoIP and other multi-media communication over IP,
in Section~\ref{sec:reqs} and propose the method to meet them in Section~\ref{sec:Method}.
Section~\ref{sec:security} analyses the
security of the method by listing and assessing the auditable information secured by it.
Section~\ref{sec:archive} describes the implementation of a secure VoIP archive.
Conclusions and an outlook are found in Section~\ref{sec:conclusions}.
A definition of and criterion for \textit{multilateral} non-repudiation,
used in Section~\ref{sec:mutuality}, are provided in the  Appendix.
\vspace*{-10mm}
\section{Requirements for non-repudiation of conversations}\label{sec:reqs}
\vspace*{-2.5mm}
From the schematic characterisation of non-repudiation in the standards~\cite{ISO10181-4,ISO13888-1},
we focus on the secure creation of evidence for later forensic inspection.
This overlaps with the basic information security targets integrity and availability of the well-known CIA triad.
To account for the particularities of the channel, we here take a communication-theoretical approach
to derive requirements for non-repudiation.
The general characteristics of the class of electronic communication that we address
are the same for a wide media range, comprising audio, video, and multi-media.
In essence it is always a full duplex or multiplex channel operating in real time using data packets,
and we subsume communication over those under the term \textit{conversation}.
Generic requirements for the non-repudiation of conversations can be profiled for specific media,
and we sometimes exemplary allude to the case of speech and VoIP.
They are grouped around the top level protection targets  \textit{congruence} and \textit{cohesion}.
We describe the latter and devise for each a minimal set of 
specific, but application- and technology-neutral requirements.
The requirements are necessary preconditions to achieve the protection targets,
and are ordered by ascending complexity. 

\textbf{T1~Congruence.} Communication theory and linguistics 
have established that the  attributions of meanings can vary between a sender and a receiver 
of a message~\cite[Chapter~6]{Searle1999},~\cite{Austin1962}
--- a basic problem for non-repudiation.
Apart from the ambiguity of language, this implies particular problems for electronic 
communication channels and media.
For digital documents bearing electronic signatures, the 
presentation problem is addressed by invoking the
`What You See is What You Sign' (WYSIWYS~\cite{LP98-WYSIWYS}) principle.
It is often tacitly assumed that presentation environments can be
brought into agreement for sender and receiver of a signed document~\cite{Schmidt2000}.
We term this fundamental target \textit{`congruence'}.
It has special traits in the case of telephony.
Essential for non-repudiation is the receiver's understanding, which leads in analogy
to the principle `What Is Heard Is What Is Signed'.
But additionally it is indispensable to assure senders (speakers) about what precisely was
received (heard).

\textbf{R1.1~Integrity} of the data in transmission, including technical environments
for sending and receiving them. For VoIP, this is to be addressed at the level of single
RTP packets and their payloads \textit{and} of an entire conversation.

\textbf{R1.2~Treatment of losses} in the channel must enable
information of senders about actually received information.    
This is independent of methods for \textit{avoidance} 
or \textit{compensation} of losses, such as Packet Loss Concealment (PLC).
Rather it means a secure detection of losses (enabled by fulfilled R1.1), enabling  a proper handling
on the application level as well as a later (forensic) inspection.

\textbf{R1.3~User interaction} policies and their enforcement finally use fulfilled R1.1 and~1.2
to ensure congruence in the inter-personal conversation.
For electronic documents this can simply amount to prescriptions about the technical environments
in which a electronically signed document must be displayed.
Or it can be an involved scheme to guarantee the agreement of contents of documents undergoing
complex transformations~\cite{LS05,PiechalskiSchmidt2006A}\eg between data formats.
For speech, it can be realised in various ways taking into account the interactive
nature of the medium. This is elaborated on in Section~\ref{sec:operation}.

\textbf{T2~Cohesion} regards the temporal dimension of conversations.
It means in particular the protection and preservation of the sequence the
information flows in all directions of the channel.
Again this is at variance with signed documents, where
temporal sequence of communication is immaterial. 
Cohesion means to establish a complete temporal context of a conversation 
usually even \textit{in absolute time}, since the temporal reference frame
of a conversation can be meaningful.

\textbf{R2.1~Start times} of conversations must be determined and recorded.
This is analogous to the signing time of documents (the assignment of which
is a requirement for qualified signatures according to the 
EU Signature Directive).

\textbf{R2.2~Temporal sequencing} of conversations must be protected and related
to the reference time frame established by fulfilling R2.1.

\textbf{R2.3~Continual authentication} of communication devices and if possible 
even communication partners is necessary\eg to prevent hijacking.  

\textbf{R2.4~Determined break points} must allow for non-repudiation of conversations
until they are terminated intentionally or inadvertently. 

From the requirements analysis it is apparent that congruence and cohesion are
complementary but not orthogonal categories.
A specific profile for VoIP is not formulated here for brevity, but 
rather included in the development of the method below.
It is understood that additionally the known 
standard requirements for electronic signatures as declarations of will and 
for non-repudiation of electronically signed documents, which are
rooted in the theory of multi-lateral security~\cite{RPM1999}, must be taken into account.
We do not address details of user authentication, 
consent to recording, general privacy, confidentiality, 
and interaction with respect to the signing as a declaration of will proper.
Nonetheless, the method proposed below enables the secure recording 
and archiving to preserve the probative value of a conversation, as
demonstrated in Section~\ref{sec:archive}.
\vspace*{-4mm}
\section{The method}\label{sec:Method}
\vspace*{-2mm}
The requirements (R2.4) entail that signing a entire conversation with a single
RSA signature by $A$ is not viable, since this yields full disposal to determine
(maliciously) the end time of signing of a conversation, and deprives $B$
of any possibility to control and verify this \textit{during conversation}. 
The opposite approach to secure single packets does not assure
cohesion (R2.2 in conjunction with R1.1), since
single RTP-packets contain only little audio data which may then easily be reordered.
Apart from that, it would be computationally expensive.
This is the prime motivation for the method we now present in general for the case of a 
bilateral conversation between $A$ and $B$, using\eg the SIP/RTP protocol combination~\cite{SIP,RTP}.
In a basic model $A$ secures the conversation as an unilateral declaration of will.
We proceed in a bottom-up fashion from the base concept of
intervals of VoIP data, over securing their integrity by
a cryptographic chain, to coping with inevitable packet loss.
For later reference we call the 
technique presented in~\ref{sec:intervals}---~\ref{sec:mutuality} below
the \textit{interval-chaining} method.
\vspace*{-2mm}
\subsection{Building intervals}\label{sec:intervals}
\vspace*{-2mm}
\textit{Intervals} are the logical units on which the protection method operates.
Intervals span certain amounts, which may be nil, of RTP packets 
for only one direction.
As bi-directional communication needs formation of intervals for both directions,
$A$ and $B$ hold buffers for packets both sent and received.
Since directions are handled differently 
w.r.t.\ packet loss, as described in Section~\ref{sec:packet-loss}, directionally homogeneous
intervals are advantageous from a protocol design viewpoint.
To resolve the full duplex audio stream into an interval sequence
we determine that intervals in the directions from and to $A$ alternate.
Intervals are enumerated as $I_{2k-1}$, $I_{2k}$, $k=1,\ldots,N$ for directions
$A\to B$ and $B\to A$, respectively.
Interval $I_l$ comprises RTP packets $(p_{l,j})$, $j=1,\ldots,K_l$,
sent or received by $A$. For the moment we assume 
that there is no packet loss.

The length of an interval (in appropriate units) is a main adjustable parameter, 
and an important degree of freedom.
Adjustable sizes of\eg data frames are not very common in communication technology,
but recent proposals~\cite{Choi2006} 
show that they can be advantageous in certain situations, like the present one.
We determine that interval boundaries are triggered by the elapse of a certain time, 
called \textit{interval duration} and denoted by $D$.
If $T$ is the duration of the conversation then $N=\lceil T/D \rceil$.
Basing intervals on time necessitates
the formation of intervals without voice data payload when a silence period exceeds $D$.
This design choice entails some signalling, transport, and cryptographic overhead.
This is however outweighed by some favourable properties.
In particular, the maximum buffer length is known from the outset,
and control of the interval duration is a direct means to cope with the (known)
slowness of (public key) cryptographic soft- and hardware.
Adjustment of $D$ therefore allows for an, even dynamical, 
trade-off between security and performance, as it controls the ratio of security data to payload data.
The alternative of triggering intervals by full-run of packet buffers
at both sides causes concurrency problems.

Since the communication channel is fully duplex, 
the sequence of intervals does not reflect the temporal sequence of audio data,
rather $I_{2k-1}$ and $I_{2k}$ comprise approximately concurrent
data sent in both directions.
But this is immaterial since intervals are only logical units and security
data for intervals can be stored separately from the RTP streams.
This is a key feature of our method. It does not affect the VoIP communication
at all but can be run in complete --- logical and even physical (extra hardware) 
--- separation from it. VoIP communication is therefore not impeded by our method.
\vspace*{-3mm}
\subsection{Cryptographic chaining}\label{sec:chaining}
\vspace*{-2mm}
The basic idea is to cryptographically secure the payload contained in each interval
and include the generated security data in the subsequent interval to form a
cryptographic chain.
We use the shorthand $(\cdot )_X\DEF \operatorname{Priv}_X(h(\cdot))$ for entity X' digital signature by
applying a private key $\operatorname{Priv}_X$ and a hash algorithm $h(\cdot )$.
$TS$ is a time-stamping authority.	
The notation $\longrightarrow$ signifies the
sending of some data.
To sign a conversation $A$ performs the following operations.
\vspace*{-1.5mm}
\begin{align*}
\text{Sec}_I  \colon & 				
M_I\DEF(D, \textsf{SIP\_Data}, \textsf{Auth\_Data}, \textsf{nonce},\ldots )\longrightarrow B;\\
&  S_0 \DEF \bigl( ( M_I )_A \bigr)_{TS} \longrightarrow B; \\			
\text{Sec}_l \colon & 
S_l \DEF ( I_l, S_{l-1} )_A \longrightarrow B; \quad l=1,\ldots,2N\\
\text{Sec}_F  \colon & 
M_F\DEF(\textsf{termination\_condition}, \ldots )\longrightarrow B;\\
&  S_F \DEF \bigl( ( M_F, S_{2N} )_A \bigr)_{TS} \longrightarrow B;
\end{align*}
\vspace*{-6mm}

\noindent In the initial step $\text{Sec}_I$, $(\cdot)_{TS}$ means a time-stamp 
applied by $TS$\eg according to RFC 3161~\cite{TSP},
and is enveloping the meta-data $M_I$ signed by $A$~(R2.1).
This may include some authentication data \textsf{Auth\_Data}
\eg $A$'s digital certificates.
To provide a broad audit trail for later inspection, 
data from the call negotiation and connection establishment,
here subsumed under \textsf{SIP\_Data}, should be included. 
The final time-stamp can be used optionally to detect drift,
and narrows down the conversation in time.
Since this is sufficient to secure the temporal context required for cohesion,
the application of time-stamps in every step, which may be costly, is not proposed.
A nonce is included in $M_I$ to prevent replay attacks.
By including $S_{l-1}$ in the signed data $S_l$ and $S_{2N-1}$ in $S_F$, and
alternation of interval directions, R1.1 and R2.2 are satisfied.		
Signatures of $A$ and additional authentication data in $M_I$ support R2.3.
If communication breaks inadvertently, 
interval chaining is verifiable up to the last interval,
thus R2.4 is satisfied, with a loss of at most one interval 
duration of conversation at its end.
$A$ controls interval timing and the operations $\text{Sec}_I$, 
$\text{Sec}_l$, and $\text{Sec}_F$ occur at times $0$, 
$\lfloor l/2 \rfloor\cdot D$, and $N\cdot D$, respectively.
\vspace*{-2mm}
\subsection{Treatment of packet loss}\label{sec:packet-loss}
\vspace*{-2mm}
Digital voice communication offers a rather high reliability 
leading generally to a higher understandability of VoIP communication 
in comparison with all predecessors. 
However, packet loss may occur and must be treated as explained in R1.2.
Denote by $\delta_l\subset\{1,\ldots, K_l\}$ the sequence of identifiers of packets actually
received by $A$ respectively $B$.
Intervals are reduced accordingly to $I_l'\DEF(p_{l,j})_{j\in\delta_l}$.
The steps $\text{Sec}_l$ are modified by a protocol to report received packages.
\vspace*{-2mm}
\begin{align*}
\text{Sec}_{2k-1}' \colon & 
\textsf{repeat} \\
& \quad \textsf{repeat} \\
& \quad\quad \textsf{interval\_termination} \longrightarrow B; \\
& \quad \textsf{until } \delta_{2k-1}\longrightarrow A;\\
& \textsf{until}\ S_{2k-1} \DEF ( I_{2k-1}', S_{2k-2} )_A \longrightarrow B;\\
\text{Sec}_{2k}' \colon & 
\textsf{repeat} \\
& \quad S_{2k} \DEF ( I_{2k}', S_{2k-1} )_A \longrightarrow B;\\
& \quad \delta_{2k}\longrightarrow B;\\
& \textsf{until success};
\end{align*}
\vspace*{-6mm}

\noindent This accounts for losses in the VoIP (RTP) channel as well as failures in the channel
for transmission of signing data.
The loop conditions can be evaluated by explicit ($\text{Sec}_{2k}'$) 
or implicit ($\text{Sec}_{2k-1}'$) acknowledgements by receivers.
\vspace*{-2mm}
\subsection{Extension to multilateral conversations}\label{sec:mutuality}
\vspace*{-2mm}
Here we present the simplest way to extend the method above to 
conference-like situations.
Multilateral non-repudiation means mutual agreement about the contents
of a conversation between all parties as defined in the Appendix.  
For implementing it for $M$ participants $A_0,\ldots,A_{M-1}$  
a round-robin scheme \cite{Shreedhar1996}
can be used to produce the required chain of signatures as in Lemma~1.
Round-robin is a simple algorithm to distribute the required security data 
between the participants of the conference.
Other base algorithms of distributed systems like flooding, echo, or broadcast
might be used, depending, for instance, on the particular topology of the
conference network.
During the round, a token is passed from participant to participant, signalling the signer role. 
If participant $A_m$ carries the token, he waits for time $D$ and buffers packets sent by himself. 
When $A_m$ terminates the interval a signalling and signing protocol is processed, which,
in contrast to the scheme above, only concerns data \textit{sent} by $A_m$. 
The numbering of intervals is as follows.
In the time span from $0$ to $D$ the packets $(p_{m;j})$ sent by $A_m$
are in the interval $I_m$. The packets emitted by $A_m$ during $[D,2D]$ are
in $I_{M+m}$, and so on.
It is here not feasible to sign merely the packets received by \textit{everyone}, 
because cumulative packet loss could be too high. 
Instead, an additional hashing indirection is included and hashes 
$H_k^\theta \DEF (h(p_{k;j}))_{j\in\theta}$ of all packets $\theta$ received 
by at least one person from $A_m$ in interval $k$ 
are distributed and can be used to check the signature
in spite of packet loss.
Let $\delta_{k}^\sigma$ denote the list of packets sent by $A_m$ and received by $A_\sigma$
in interval $k$.
Set $R_m\DEF \{0,..,M-1\}\backslash m$ and let
$r\geq 0$ be the round number.
In order to account for latencies in reporting of packet loss, computing hashes,
and signing, we introduce a parallel offset in the round-robin scheme.
In round $r$ participant $A_m$ carrying the token terminates interval with number
$\widehat{k}(r,m)\DEF rM^2+(M+1)m+1$.
He secures the set of intervals $\widehat{I}(r,m)\DEF\bigl(\widehat{k}(r,m)-M\cdot\{0,\ldots,M-1\}\bigr)\cap\NN$.
\vspace*{-2mm}
\begin{align*}
\text{Sec\_mult}_{r,m} \colon &\forall \sigma \in R_m \textsf{ do}\\
 &\quad \textsf{repeat} \\
 &\quad \quad \textsf{interval\_termination} \longrightarrow A_\sigma;\\
 &\quad \textsf{until}\ (\delta_k^\sigma)_{k \in \widehat{I}(r,m) } \longrightarrow \ A_m;\\
& \textsf{od};\\
& \theta_{k} \DEF \cup_{\sigma\in R_m} \delta_{k}^\sigma \textsf{ for } k\in\widehat{I}(r,m);\\
& D_{r,m} \DEF \bigl((\delta_{k}^\sigma)_{\sigma \in R_m}, H_{k}^{\theta_{k}}\bigr)_{k \in \widehat{I}(r,m)};\\
& S_{r,m} \DEF (D_{r,m}, S_{\operatorname{pred}(r,m)})_{A_m};\\
&\forall \sigma' \in R_m \textsf{ do}\\
 &\quad\textsf{repeat}\\
 &\quad\quad (S_{r,m},D_{r,m}) \longrightarrow A_{\sigma'};\\
 &\quad\textsf{until success};\\
& \textsf{od};
\end{align*}
\vspace*{-6mm}

\noindent 
The preceding security value $S_{\operatorname{pred}(r,m)}$ bears indices
\vspace*{-2mm}
\[
\operatorname{pred}(r,m)=
\begin{cases}
  (r,m-1) & \text{if } m\geq1; \\
  (r-1,M-1) & \text{if } r\geq1,\ m=0;\\
  I & \text{otherwise,}
\end{cases}
\]
\vspace*{-4mm}

\noindent 
where $I$ stands for the initialisation interval which can be constructed as in the preceding sections,
replacing single sending by broadcast with acknowledgements.
The numbering scheme for Intervals and the evolving sequence of security values is shown in
Figure~\ref{fig:multisig_scheme} below.
In effect, $A_m$ broadcasts (with acknowledgement) a signature over hashes of all packets received 
by at least one other participant. 
This is the \textit{common} security data with which the chain can be continued. 
According to Lemma~1, non-repudiation of the total, multilateral conversation for 
the first interval duration from time $0$ to $D$ is achieved after execution of
$\text{Sec\_mult}_{2,M-1}$ at time $2M\cdot D$.
With each further execution of $\text{Sec\_mult}$ a subsequent piece of conversation
of length $D$ obtains multilateral non-repudiation.
\begin{figure}[tbp]
\centering
\includegraphics[scale=0.5]{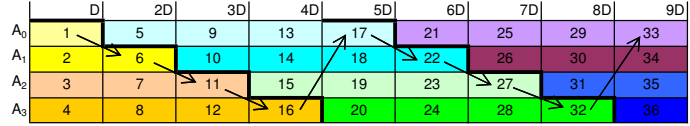}
\caption{Numbering of intervals in the case of $4$ participants along the time axis.
Arrows indicate the sequence of security values $S$. Thicker borders separate rounds.
Equally coloured intervals are secured in a single operation $\text{Sec\_mult}_{r,m}$.}
\label{fig:multisig_scheme}
\vspace*{-4mm}
\end{figure}

In case of call termination, $2M+1$ finalisation steps without audio data 
(two final rounds plus finishing by the participant carrying the token at the time of termination) 
are required to obtain non-repudiation of the last interval in time.
Joining and leaving a signed multilateral call while the signature is created by the participants
can be enabled through finalisation. 
If participant $B$ requests
to join the call, $A_m$, who posses the token, initiates a finalisation and $B$
can join after this (inserted as $m+1$). 
In the case that a participant likes
to leave he awaits the token and finalises including a $\textsf{leave}$ message.
\vspace*{-2mm}
\subsection{Operational policies}\label{sec:operation}
\vspace*{-2mm}
We do not lay out a complete set of rules for the operation of a system
using the non-repudiation method above. 
Rather we list the most obvious ones
and stress the most important point of monitoring  and treatment of packet loss, or rather understandability.

To account for requirement R1.3, users should be signalled at any time during a conversation
about the signature status of it.
This necessitates to an extent specified by application-specific policies the cryptographic 
verification of the interval chaining, and continual evaluation of relevant information,
see Section~\ref{sec:forensics}.
Additionally a \textit{secure voice signing terminal} 
should control every aspect of user interaction and data transmission.
This is elucidated in~\cite{WWRF17}.

To maintain congruence and 
mitigate attacks aiming at mutilating a conversation, 
 packet loss and the ensuing level of understandability
must  permanently be monitored. 
When the packet loss is above a configurable threshold,
an action should be triggered according to determined policies.
The principle possibilities are:
1.~ignore;
2.~notify users while continuing signing;
3.~abort the signing; and
4.~terminate call. 
The first two options open the path for attacks. 
Termination of the call is the option for maximum security. 
From a practical viewpoint,
the loss threshold is seldom reached without breakdown of the connection 
anyway due to insufficient understandability or timeouts.

Options~3 and~4 provide a `Sollbruchstelle' (predetermined break point) 
for the probative value of the conversation. 
In contrast, most other schemes for securing the integrity of streamed data\eg 
the signing method of~\cite{Perrig2000} aim at loss-tolerance, 
for instance allowing for the verification of the stream signature with some probability 
in the presence of packet loss. 
We suggest that for the probative value of conversations, the former is advantageous. 
A signed call with an intermediate  gap can  give rise to speculations over 
alternatives to fill it, which are restricted by syntax and grammar, 
but can lead to different semantics. 
Using this, a clever and manipulative attacker could delete parts of the communication 
to claim with certain credibility that the remnants have another meaning than intended by 
the communication partner(s). 		
If the contents of a conversation after such an intentional deletion are 
unverifiable and thus cannot be used to prove anything, this kind of attack is effectively impeded.
\vspace*{-5mm}
\section{Security considerations}\label{sec:security}
\vspace*{-2mm}
We corroborate the statement that interval chaining 
can achieve non-repudiation for VoIP conversations,
based on the information generally secured by interval chaining. 
An analysis based on an
instance of a system architecture (the VoIP archive presented in Section~\ref{sec:archive} below) 
and possible attacks is contained in~\cite{HKS06A}.
\vspace*{-3mm}
\subsection{Auditable information}\label{sec:forensics}
\vspace*{-2mm}
In this section we analyse the information that can be gained and proved to have 
integrity in a call secured by interval chaining.
Table~\ref{tab:Checks} gives a, perhaps incomplete, overview over this audit data, which may be
amenable to forensic inspection\eg by an expert witness in court, or, on the other extreme,
applicable during the ongoing conversation, or both.
\begin{table}[htb]
	\centering
\begin{footnotesize}
		\begin{tabularx}{125mm}{|p{115pt}|p{29pt}|p{50pt}|X|p{45pt}|}		
\hline
Auditable item & Req. & Protection target & Verifies/indicates & 	When applicable\\
\hline 
\hline
Initial time stamp	& 2.1 & Cohesion	& Start time	& Always			\\
\hline
Initial signature \& certificate & 2.3 & Cohesion &	Identity of signer &	Always	\\
\hline
Interval Chaining	& 2.2, 1.1  &		Cohesion &		Interval integr. \& order &		Always \\
\hline
Packet loss in intervals	& 1.2, 2.4  &	Congruence &	QoS, understandability &		Always \\
\hline
Monotonic increase of RTP-sequence numbers		& 1.1,  2.2 &	Integrity \& cohesion &		RTP-stream plausibility	 &	Always \\
\hline
Relative drift of RTP-time marks against system time	& 	2.2  &		Cohesion	 & RTP-stream plausibility &	During convers. \\
\hline
Relative drift of RTP-time marks against $\lfloor l/2 \rfloor\cdot D$	& 	2.2  &		Cohesion &		Packet \& stream plausibility	 &	Ex post \\
\hline
No  overlaps of RTP-time marks at interval boundaries & 	2.2  &		Cohesion &	Packet \& stream plausibility	 &	Always \\
\hline
Replay-window	 &	&	Integrity  &		Uniqueness of recorded audio stream &		Always \\
\hline
Final time stamp	& 	2.2  &		Cohesion &		Conversation duration	 &	Ex post \\
\hline
Forensic analysis of recorded conversation	 & &	(Semantic) authenticity &		Speaker identity, mood, lying, stress,  etc. &	Ex post, fo­rensic \\
			\hline
		\end{tabularx}\end{footnotesize}\\
\caption{Auditable information of a conversation secured with interval chaining.
Columns: Secured data item audited, Non-repudiation requirement addressed, Protection target supported, 
Actual information indicated or verified, and when is the check applicable.}
	\label{tab:Checks}
\vspace*{-7mm}
\end{table}%
\vspace*{-3mm}
\subsection{Comparison with SRTP and IPsec}\label{sec:IPSEC}
\vspace*{-2mm}
The well-known security methods SRTP and IPsec
address the protection of confidentiality,
authenticity and data integrity on the application, respectively network layer, and can be 
applied to VoIP and as well in parallel with interval-chaining.
We want to show salient features of interval-chaining, which distinguishes it from both
standards and in our view provides a higher level of non-repudiation and even practicality.
On the fundamental level, both SRTP and IPsec necessarily operate on the packet level and
do not by themselves provide protection of the temporal sequence and cohesion of a 
VoIP conversation.
While it is true that pertinent information can be reconstructed from the 
RTP sequence numbers, in turn protected by hash values, such an approach would have some
weaknesses, which taken together do not allow full non-repudiation.
In particular, RTP sequence numbers can suffer from roll-overs and though their integrity
is secured in transmission, they can still be rather easily be forged by the sender,
since they belong to protocol stacks which are not especially secured in common systems.
While packet loss can be detected or reconstructed using sequence numbers, interval chaining
yields a well-defined, tunable, \textit{and cryptographically secured} means to deal with
it during an ongoing conversation, significantly limiting potential attack vectors.
In essence, RTP sequence numbers are not designed to ensure a conversation's 
integrity and thus have lower evidentiary value in comparison to chained intervals.
From the viewpoint of electronic signatures, their level of message, respectively, 
conversation authentication can only be achieved with an protocol-independent means to 
manage authentication data such as asymmetric keys\ie a Public Key Infrastructure.
The connection and session dependent key handling of IPsec and SRTP, relying on HMACs and 
merely allowing for symmetric keys deprived of authentication semantics, are generally insufficient 
for non-repudiation.
Interval chaining is an independent means to control the cryptographic workload benefiting
scalability.
Finally, NAT traversal is a problem for network layer integrity protection like 
IPsec since rewriting  IP headers invalidates corresponding
hash values (a solution has been proposed by TISPAN~\cite{TISPAN}).
This problem does not occur with the interval chaining method,
since only RTP headers, not IP headers of packets need to be 
(and are in the implementation below) signed.
\vspace*{-4mm}
\section{Application to a secure VoIP archive}\label{sec:archive}
\begin{figure}[hbtp]
\vspace*{-4mm}
\centering
\subfigure[]{%
\label{subfig:sysmodel}\includegraphics[scale=0.42]{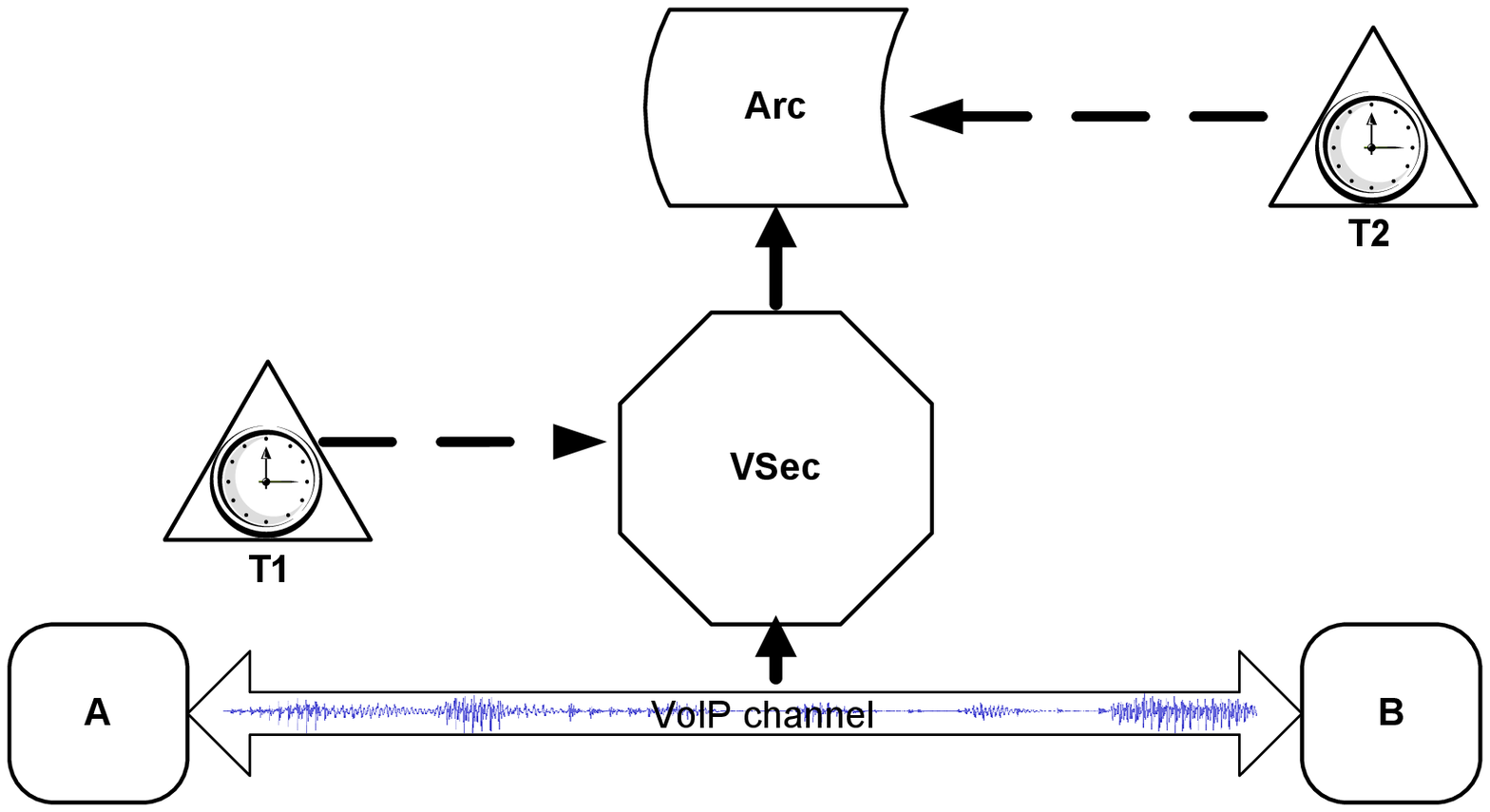}}%
\hfill%
\subfigure[]{%
\label{subfig:checks}\includegraphics[scale=0.36]{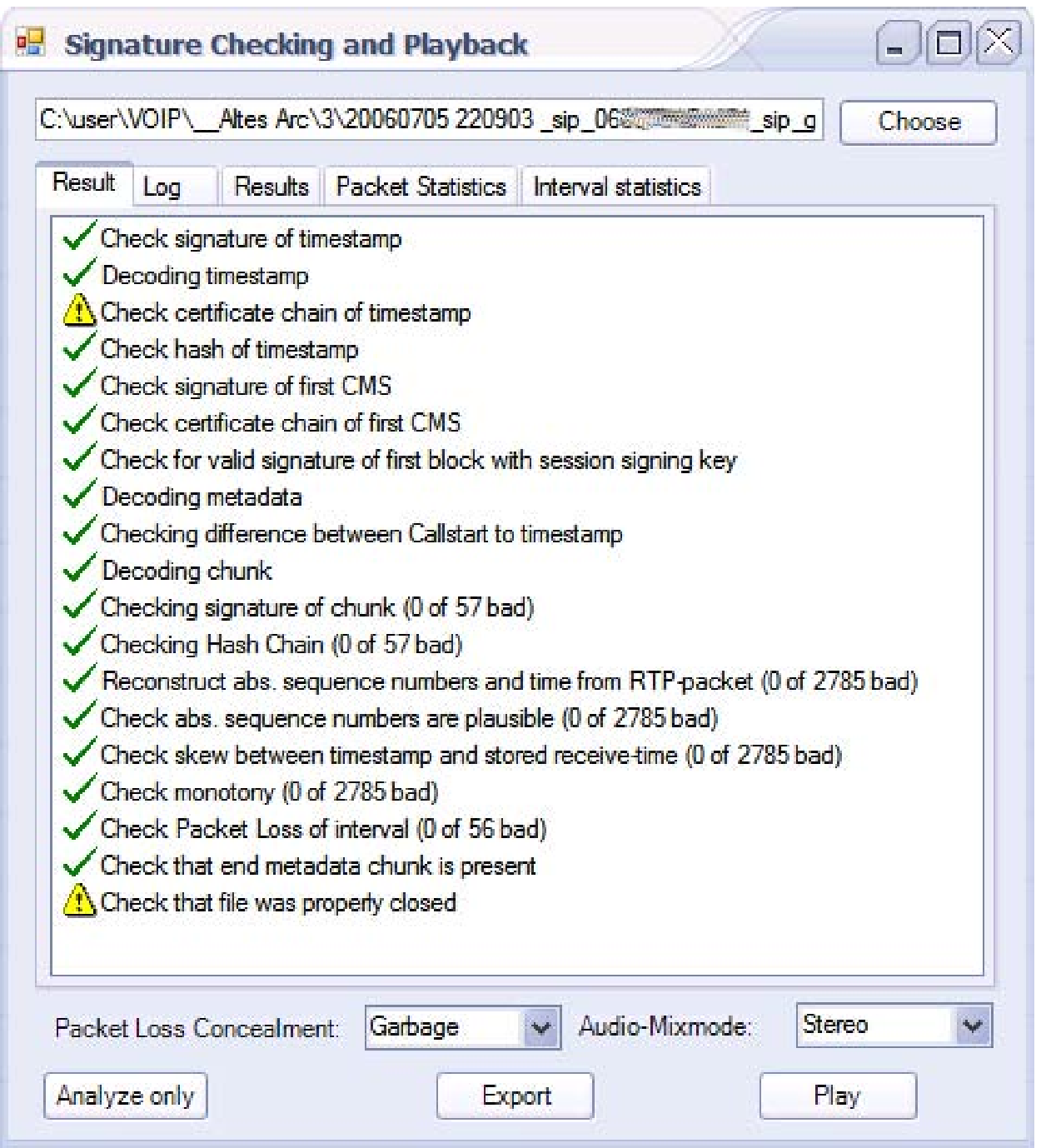}}
\caption{(a) High-level modular architecture of a secure voice archive. 
(b) The verification and playback tool performing checks on the archived call.}
\vspace*{-4mm}
\end{figure}

\vspace*{-2mm}
In this section we present an efficient self-signed archive for VoIP calls and its system architecture.
It was implemented as a prototype together with a verification and playback tool, requires
no modification to the terminal equipment, and secures the ongoing conversations `on the fly'.
Section 5.1~was partially published in~\cite{HKS06A}.
It uses timestamps to secure the exact starting-time of a conversation and not the moment of archiving.
The main design principle is
that of minimal technical requirements on the communication clients.
Figure~\ref{subfig:sysmodel} shows the communication between two partners \CA and \CB over
a VoIP channel. 
The security component \VS, the component implementing the interval chaining, 
can listen to the communication at any point in the channel. 
Neither the exact position in the channel nor the technical
method by which \VS intercepts it is essential for the architecture and its security
properties.
It can be located at the site of either
of the parties \CA or \CB\eg in the case of call-centre applications. 
The channel is \textit{not} required to be digital, let alone SIP/RTP based, end-to-end, 
provided that there is some part of the channel which is VoIP. 
This condition is already met in many mobile 
and public switched networks.  
Accordingly, the phones used by \CA and \CB need not be ISDN or VoIP phones. 
\VS will often be under the control of one of the parties or even be integrated in their
VoIP infrastructure. 

The role of \VS can be passive or dual, for instance to enforce
policies, cf.~Section~\ref{sec:operation}. 
The separation of such a component is standard in 
security engineering where it is
commonly known as a reference monitor~\cite{Anderson1972}.
The component \Arc denotes the archive to which the secured VoIP communication is submitted and
then persistently stored, with due consideration of long-term security.
\TTA and \TTB are time-stamping authorities which come into play to raise
resilience against attacks\eg if \VS, \Arc or both are compromised.
\vspace*{-2mm}
\subsection{Implementation}\label{sec:archive-details}
\vspace*{-2mm}
The archive system has been implemented as a prototype and tested with several
soft phones and devices\eg AVM's Fritz!Box~\cite{Fritz} using SIP/RTP.
For \CB we used mobile phones, ISDN phones, and also SIP software clients.
\VS was implemented using C\#, running on an embedded x86 based PC without keyboard,
mouse or video ports on Linux OS with the Mono-framework. 
It was placed as an outbound proxy, substituting $A$'s original one, 
between \CA and the Internet using its two NICs, thus supporting multiple concurrent clients and calls.
The proxy modifies RTP ports and IP addresses contained in the SIP packets redirecting 
them to itself and forwards them to the original recipients.
A traditional PC was used for \Arc, connected over
a reliable TCP channel (TLS can be used for privacy) using the third NIC of \VS .

Intervals are formed as in Section~\ref{sec:chaining} and 
\ref{sec:packet-loss}, 
though packet loss is here handled by simply storing a list of 
the received packets. 
This is an obvious necessity as the terminal equipment cannot
be modified in a pure archiving scenario.
\VS carries an X.509 certificate together with the 
private (RSA) key to sign intervals, including $S_0$ and $S_F$
containing meta data. 
The signing operation $(\cdot )_A$ is extended beyond the presented
formal scheme by not using\eg RSA signatures directly, but instead PKCS\#7 signed envelope (where
data is to be stored together with its signature) and PKCS\#7 detached signatures (where
signatures are stored alone). 
This has the advantage to store and transport certificates
and certificate chains in combination with the data. 
For storage efficiency, only the first interval's PKCS\#7 signed envelope container 
contains the whole certificate chain up to, but not including the root, 
while all other containers do not need to carry this redundant 
information. 
$S_0$ is also wrapped in a signature from the 
time stamp service \TTA.

The certificate and private key of \VS are not only used to sign the conversation, 
but also  to authenticate \VS toward \Arc. 
Each completed interval is immediately transmitted to \Arc, which 
then performs several tests, see Table~\ref{tab:Checks}, including 
signature verification. 
It then stores it as chunks into an open file. 
This yields resilience even against severe attacks\eg a compromised \VS. 
\Arc checks the trusted third party time stamp from \TTA immediately at the start of the call, 
thus the live archiving cannot be delayed, and also checks each interval as soon as 
it arrives. 
Therefore, a compromised \VS would have
to forge conversations in duration of an interval\ie in real time, which is infeasible.
On the other hand, a compromised \Arc does not posses the certificate. 
Even if both collude, 
they were still be bound inside the two timestamps from \TTA and \TTB, which could be compared\eg to an
itemised bill from the phone company.

The implemented system performed very well and was able to archive multiple parallel calls on the mentioned
hardware without notable load or memory pressure. 
The duration of an interval is one of the main configuration
parameters to be tuned. One second proved to be sufficient to provide a high 
level of security for the context of the talk on the one hand, and
on the other hand to keep the computational power required
by far low enough for the used processor, and also the storage overhead 
(400 bytes for PKCS\#7 signatures without embedded certificates)
to payload ratio small. 
\vspace*{-4mm}
\section{Conclusions}\label{sec:conclusions}
\vspace*{-2mm}
IP-based multimedia communication is not restricted to VoIP, for instance 
by now, several video conferencing systems are maturing, some of which
are based on sophisticated peer-to-peer communication~\cite{Zuhlke2004,Popovici2003}.
Moreover the service quality and availability of the new communication channels
is constantly increasing through developments like packet loss
concealment (PLC) for audio~\cite{Perkins1998} and even video~\cite{zhu:52,Lee2006} streams.
Our proposed method for non-repudiation is applicable in all these contexts.
Its adoption would pave the way for a new paradigm for trustworthy, inter-personal communication.

The next steps in our research are to i) implement the operational context for
electronic signatures over speech\ie user interaction and signalling,
ii) devise a trustworthy signature terminal for that purpose, preferably using
Trusted Computing technology on mobile devices\eg to secure audio I/O and
processing, iii) extend the method to conferences and other media than VoIP.
\vspace*{-4mm}
\section*{Appendix: Multilateral non-repudiation with signatures}\vspace*{-2mm}
Contracts embodied in paper or electronic form often bear the attribute of mutual non-repudiation
(in the case of two parties) if both $A$ and $B$ receive a copy of the doubly signed document,
and know with certainty that the other party has it as well. 
To achieve mutuality in digital contexts, some kind of bi-lateral electronic signing must be performed.
For electronic documents, it is well known that a simple, three-step protocol (called BAKO~\cite{BAKO-Ur}),
enveloping three signatures in the sequence $ABA$ around the document, suffices.

This section superficially formalises the notion of a multilateral agreement
between parties $A_1,\ldots,A_M$ on the fact \textsf{Comm} which is the logical 
assertion that a certain communication has taken place between them.
A more rigorous treatment in terms of formal languages can be found in~\cite{GO01}.
Assume that \textsf{Comm} can be acknowledged with certainty by $A_\ell$\ie
$A_\ell$ asserts his knowledge that no sequence of events can occur which negates \textsf{Comm}, by
digitally signing it\ie by forming $(\textsf{Comm})_{A_\ell}$.
We write
\[
(\textsf{Comm})_{A_\ell}\Longrightarrow\textsf{Knows}_{A_\ell}(\textsf{Comm}).
\]
\textit{Multilateral non-repudiation} is partial \textit{common knowledge} of \textsf{Comm} in the sense
that every party knows that every party knows $\textsf{Comm}$\ie
\[
\bigwedge\nolimits_{\ell=1,.\ldots,M}
\textsf{Knows}_{A_\ell}
\left(\bigwedge\nolimits_{k=1,.\ldots,M}
\textsf{Knows}_{A_k}(\textsf{Comm})\right).
\]
In extension of BAKO for mutuality we find
\begin{lemma}
  Multilateral non-repudiation of $\textnormal{\textsf{Comm}}$ is achieved for $A_1,\ldots,A_M$ by formation of
\[
  (\ldots(((\ldots((\textnormal{\textsf{Comm}})_{A_1})_{A_2}\ldots)_{A_M})_{A_1})_{A_2}\ldots)_{A_M}.
\]
\end{lemma}
\begin{proof}
  Replacing $B$ in the BAKO scheme $ABA$ by $A_2\ldots A_M$ we see that
formation of $((\ldots((\textnormal{\textsf{Comm}})_{A_1})_{A_2}\ldots)_{A_M})_{A_1}$ establishes
\[
\textsf{Knows}_{A_1}
\left(\bigwedge\nolimits_{k=1,.\ldots,M}
\textsf{Knows}_{A_k}(\textsf{Comm})\right).
\]
The statement follows by iteration.
\end{proof}
\providecommand{\noopsort}[1]{} \providecommand{\singleletter}[1]{#1}


\vspace*{-6mm}
\begin{thebibliography}{10}
\vspace*{-2mm}

\bibitem{VoIP-Study}
Kavanagh, J.: 
Voice over IP special report: From dial to click. 
\url{http://www.computerweekly.com/Articles/2006/02/14/214129/VoiceoverIPspecialreportFromdialtoclick.html}, visited 1.3.2006.

\bibitem{SRTP}
Baugher, M., et al.: The Secure Real-time Transport Protocol (SRTP). 
RFC 3711, IETF, March 2004. \url{http://www.ietf.org/rfc/rfc3711.txt}

\bibitem{Edison1911}
Edison, T.A.:
\newblock Recording-telephone.
\newblock United States Patent P.No.:1,012,250, United States Patent Office,
  Washington, DC (1911) Patented Dec. 19, 1911.

\bibitem{Merkle1989}
Merkle, R.C.:
\newblock A certified digital signature.
\newblock In Brassard, G., ed.: Advances in Cryptology (CRYPTO '89). Number 435
  in LNCS, Springer-Verlag (1989)  218--238 Republication of the 1979 original.

\bibitem{Strasser-Thesis2001}
Strasser, M.:
\newblock M{ö}glichkeiten zur {G}estaltung verbindlicher {T}elekooperation.
\newblock Master's thesis, Universit{ä}t Freiburg, Institut f{ü}r
  Informatik und Gesellschaft (2001)

\bibitem{KKKZ2000}
Kabatnik, M., Keck, D.O., M.~Kreutzer, A.Z.:
\newblock Multilateral security in intelligent networks.
\newblock In: Proceedings of the IEEE Intelligent Network Workshop. (2000)
  59--65

\bibitem{noiserobustspeakeraut}
Poh, N., Bengio, S.:
\newblock {Noise-Robust Multi-Stream Fusion for Text-Independent Speaker
  Authentication}.
\newblock In: Proceedings of The Speaker and Language Recognition Workshop
  (Odyssey). (2004)

\bibitem{fusionspeakeraut}
Rodriguez-Linares, L., Garcia-Mateo, C.:
\newblock {Application of fusion techniques to speaker authentication over IP
  networks}.
\newblock IEEE Proceedings-Vision Image and Signal Processing \textbf{150}
  (2003)  377--382

\bibitem{Hollien2001}
Hollien, H.:
\newblock Forensic Voice Identification.
\newblock Academic Press, London (2001)

\bibitem{Goodwin1981}
Goodwin, C.:
\newblock Conversational organization: Interaction between speakers and
  hearers.
\newblock Academic Press, New York (1981)

\bibitem{LP98-WYSIWYS}
Landrock, P., Pedersen, T.:
\newblock {WYSIWYS?} What You See Is What You Sign?
\newblock Information Security Technical Report, \textbf{3} (1998)  55--61

\bibitem{ISO10181-4}
ISO:
\newblock Information {T}echnology: {S}ecurity {F}rameworks for {O}pen
  {S}ystems: {N}on-{R}epudiation {F}ramework.
\newblock Technical Report ISO10181-4, ISO (1997)

\bibitem{ISO13888-1}
ISO:
\newblock Information {T}echnology: {S}ecurity {T}echniques - {N}on
  {R}epudiation - {P}art 1: {G}eneral.
\newblock Technical Report ISO13888-1, ISO (1997)

\bibitem{Searle1999}
Searle, J.R.:
\newblock Mind, {L}anguage and {S}ociety.
\newblock Basic Books, New York (1999)

\bibitem{Austin1962}
Austin, J.L.:
\newblock How to {D}o {T}hings with {W}ords.
\newblock Harvard University Press, Cambridge, Mass. (1962)

\bibitem{Schmidt2000}
Schmidt, A.U.:
\newblock {Signiertes XML und das Pr{ä}sentationsproblem}.
\newblock Datenschutz und Datensicherheit \textbf{24} (2000)  153--158

\bibitem{LS05}
Schmidt, A.U., Loebl, Z.:
\newblock Legal security for transformations of signed documents: Fundamental
  concepts.
\newblock In Chadwick, D., Zhao, G., eds.: EuroPKI 2005. Volume 3545 of Lecture
  Notes in Computer Science., Springer-Verlag (2005)  255--270

\bibitem{PiechalskiSchmidt2006A}
Piechalski, J., Schmidt, A.U.:
\newblock Authorised translations of electronic documents.
\newblock In Venter, H.S., Eloff, J.H.P., Labuschagne, L., Eloff, M.M., eds.:
  Proceedings of the ISSA 2006 From Insight to Foresight
  Conference, Information Security South Africa (ISSA) (2006)

\bibitem{RPM1999}
Rannenberg, K., Pfitzmann, A., M{ü}ller, G.:
\newblock {IT} {S}ecurity and {M}ultilateral {S}ecurity.
\newblock In M{ü}ller, G., Rannenberg, K., eds.: Multilateral Security in
  Communications. Volume~3 of Technology, Infrastructure, Economy.,
  Addison-Wesley (1999)  21--29

\bibitem{SIP}
Rosenberg, J., et al.: SIP: Session Initiation Protocol.
RFC 3261, IETF, June 2002. \url{http://www.ietf.org/rfc/rfc3261.txt}

\bibitem{RTP}
Schulzrinne, H., et al.: RTP: A Transport Protocol for Real-Time Applications.
RFC 1889, IETF, January 1996. \url{http://www.ietf.org/rfc/rfc1889.txt}

\bibitem{TSP}
Adams, C., et al.: Internet X.509 Public Key Infrastructure Time-Stamp Protocol (TSP).
RFC 3161, IETF, August 2001. \url{http://www.ietf.org/rfc/rfc3161.txt}

\bibitem{Choi2006}
Choi, E.C., Huh, J.D., Kim, K.S., Cho, M.H.:
\newblock Frame-size adaptive MAC protocol in high-rate wireless personal area
  networks.
\newblock ETRI Journal \textbf{28} (2006)  660--663

\bibitem{Shreedhar1996}
Shreedhar, M., Varghese, G.:
\newblock Efficient fair queuing using deficit round-robin.
\newblock IEEE/ACM Transactions on Networking \textbf{4} (1996)  375--385

\bibitem{WWRF17}
Hett, Ch., Kuntze, N.,  Schmidt, A. U.:
Security and non repudiation of Voice-over-IP conversations.
To appear in: Proceedings of the Wireless World Research Forum (WWRF17), 
Heidelberg, Germany,  15-17 November 2006.

\bibitem{Perrig2000}
Perrig, A., Tygar, J.D., Song, D., Canetti, R.:
\newblock Efficient authentication and signing of multicast streams over lossy
  channels.
\newblock In: SP '00: Proceedings of the 2000 IEEE Symposium on Security and
  Privacy, Washington, DC, USA, IEEE Computer Society (2000)  56--75

\bibitem{HKS06A}
Hett, C., Kuntze, N., Schmidt, A.U.:
\newblock A secure archive for Voice-over-IP conversations.
\newblock In et~al., D.S., ed.: To appear in the Proceedings of the 3rd Annual
  VoIP Security Workshop (VSW06), ACM (2006)
\url{http://arxiv.org/abs/cs.CR/0606032}

\bibitem{TISPAN}
Telecoms \& Internet converged Services \& Protocols for Advanced Networks (TISPAN)
\url{http://www.tispan.org/}, see also the Whitepaper
\url{http://www.newport-networks.com/cust-docs/91-IPSec-and-VoIP.pdf}

\bibitem{Anderson1972}
Anderson, J.P.:
\newblock Computer security technology planning study, Volume II.
\newblock Technical Report ESD-TR-73-51, Electronic Systems Division, Air Force
  Systems Command, Hanscom Field, Bedford, MA (1972)

\bibitem{Fritz}
 AVM Fritz!Box. \url{http://www.avm.de}

\bibitem{Zuhlke2004}
Z{ü}hlke, M., K{ö}nig, H.:
\newblock A signaling protocol for small closed dynamic multi-peer groups.
\newblock In: Proceedings of High Speed Networks and Multimedia Communications,
  7th IEEE International Conference (HSNMC 2004), Toulouse, France. Volume 3079
  of LNCS., Springer-Verlag (2004)  973--984

\bibitem{Popovici2003}
Popovici, E.C., Mahlo, R., Z{ü}hlke, M., K{ö}nig, H.:
\newblock Consistency support for a decentralized management in closed
  multiparty conferences using {SIP}.
\newblock In: ICON 2003. The 11th IEEE International Conference on Networks,
  IEEE Press (2003)  295--300

\bibitem{Perkins1998}
Perkins, C., Hodson, O., Hardman, V.:
\newblock A survey of packet loss recovery techniques for streaming audio.
\newblock IEEE Network \textbf{12} (1998)  40--48

\bibitem{zhu:52}
Zhu, Q.F., Kerofsky, L.:
\newblock Joint source coding, transport processing, and error concealment for
  {H}.323-based packet video.
\newblock In Aizawa, K., Stevenson, R.L., Zhang, Y.Q., eds.: Visual
  Communications and Image Processing '99. Volume 3653 of Proceedings of SPIE.,
  SPIE (1998)  52--62

\bibitem{Lee2006}
Lee, P.J., Lin, M.L.:
\newblock Fuzzy logic based temporal error concealment for {H.264} video.
\newblock ETRI Journal \textbf{28} (2006)  574--582

\bibitem{BAKO-Ur}
Kolletzki, S.:
\newblock Secure internet banking with privacy enhanced mail - a protocol for
  reliable exchange of secured order forms.
\newblock Computer Networks and ISDN Systems \textbf{28} (1996)  1891--1899

\bibitem{GO01}
Grimm, R., Ochsenschl{ä}ger, P.:
\newblock {B}inding {C}ooperation. {A} {F}ormal {M}odel for {E}lectronic
  {C}ommerce.
\newblock Computer Networks \textbf{37} (2001)  171--193

\end{thebibliography}
\end{document}